\documentclass[twocolumn,runningheads]{svjour2}
\smartqed  
\usepackage{graphicx}
\newcommand{\beq}{\begin{eqnarray}}
\newcommand{\eeq}{\end{eqnarray}}

\newcommand{\amx}{{$\alpha_{\mu x}$~}}

\newcommand{\nupeak}{$\nu_{peak}$~}
\journalname{Astrophysics and Space Science}
\begin{document}

\title{Blazar Duty-Cycle at $\gamma$-ray Frequecies: Constraints from
Extragalactic Background Radiation and Prospects for AGILE and GLAST
}



\author{Carlotta Pittori         \and
        Elisabetta Cavazzuti      \and 
    Sergio Colafrancesco       \and
    Paolo Giommi
}


\institute{C. Pittori \and E. Cavazzuti \and P. Giommi
        \at
              ASI Science Data Center, ASDC c/o ESRIN, via G. Galilei I-00044 Frascati, Italy
             \\
              Tel.: +39-06-94188878\\
              Fax: +39-06-94188872\\
              \email{carlotta.pittori@asdc.asi.it}            \\
          \and
           S. Colafrancesco \at
    INAF-Osservatorio Astronomico di Roma,
    via di Frascati 33, I-00040 Monteporzio, Italy
}

\date{Received: date / Accepted: date}

\maketitle

\begin{abstract}
%
We take into account the constraints from the observed extragalactic
$\gamma$-ray background to
estimate the maximum duty cycle allowed for a selected sample of
WMAP Blazars, in order to be detectable by
AGILE and GLAST $\gamma$-ray experiments.
For the nominal sensitivity values of both instruments, we identify
a subset of sources which can in principle be detectable
also in a steady state
without over-predicting the extragalactic background.
This work is based on the results of a recently derived
Blazar radio LogN-LogS obtained by combining several multi-frequency surveys.
\end{abstract}

\section{Introduction}
\label{intro}
Blazars are the dominant population of extragalactic
sour\-ces at microwave, $\gamma$-rays and TeV energies. They  represent
$5\%-8\%$ of all AGNs and are powerful sources emitting a continuum of
electromagnetic radiation from a relativistic jet viewed closely along the line
of sight. The large observed variety of Blazar Spectral Energy Distributions
(SEDs) can be reproduced, at least in first approximation, by simple
Synchrotron Self Compton (SSC) emission model, composed of a synchrotron
low-energy component that peaks between the far infrared and the X-ray band,
followed by an Inverse Compton component that has its maximum in the hard X-ray
band or at higher energies, and may extend into the $\gamma$-ray or even the
TeV band. Those Blazars where the synchrotron peak is located at low ($\sim$
infrared) energy are usually called Low energy peaked Blazars or LBL, while
those where the synchrotron component reaches the X-ray band are called High
energy peaked Blazars or HBL.
Blazars may also be subdivided in BL Lacertae types (BL Lacs
$\sim 20\%$ of all Blazars), characterized  by
strong non-thermal emission with no or very weak emission lines and
in Flat Spectrum Radio Quasars (FSRQs $\sim 80\%$)
which share the strong non-thermal emission of BL Lacs but also show intense broad line emission.
LBL sources, mostly FSRQ and few BL Lacs, are the large majority among Blazars
and are usually discovered in radio surveys,
while HBL objects all of BL Lac type, are preferentially found in X-ray flux limited surveys.

Despite the relatively low space density of Blazars,
their strong emission across the entire electromagnetic spectrum makes
them potential candidates as significant contributors to extragalactic Cosmic Backgrounds.
Giommi et al. 2005 \cite{gio06} have recently re-assesed
the Blazar contribution to the microwave (CMB), X-ray (CXB),
$\gamma$-ray (CGB) and TeV Cosmic backgrounds
based on a new estimation of the Blazar radio LogN-LogS,
assembled combining several radio and multi-fre\-quen\-cy surveys.
It was shown that Blazars add a non-thermal component to the overall 
Cosmic Background
that at low frequencies contaminates the CMB fluctuation spectrum. 
At higher energies
(E $>$100 MeV) the estimated Blazar
collective emission over-predicts the extragalactic background
by a large factor, thus implying that Blazars not only dominate the
$\gamma$-ray sky (cfr. ref.~\cite{padovani}),
but also that their average duty cycle at these 
frequencies must be rather low.

In this paper we analize a sample of
WMAP detected Blazars and we estimate
the maximum duty cycle allowed, taking into account the constraints from the
observed extragalactic $\gamma$-ray background, in order to be detectable by
AGILE and GLAST
for the nominal sensitivity values of both instruments.
%
\section{Observational Constraints and Blazar $\gamma$-ray Duty Cycle}
\label{sec:method}

\begin{figure}
\centering
\includegraphics[width=0.60\textwidth]{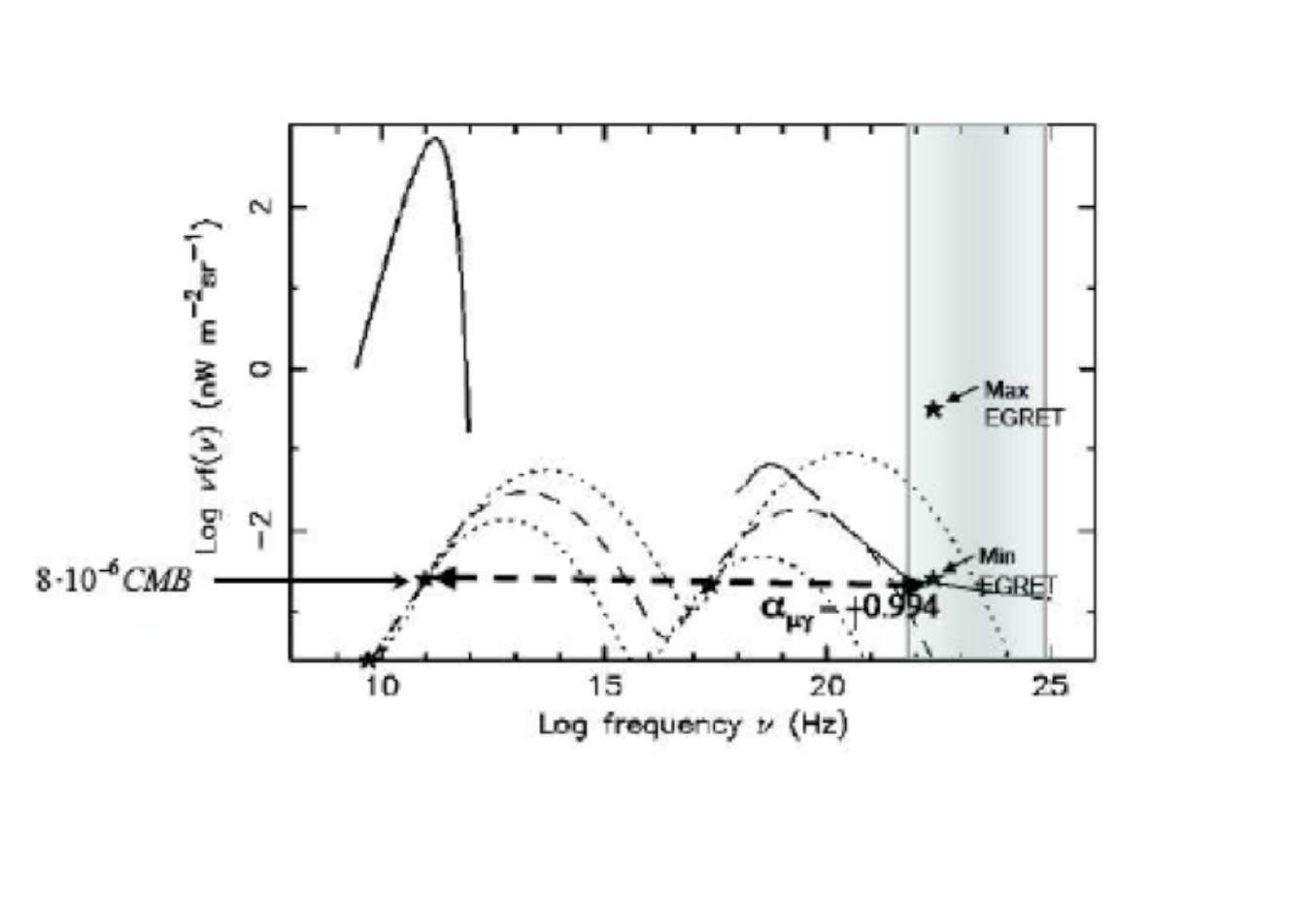}
\vskip -0.7 truecm \caption{The possible contribution of LBL Blazars to the
Hard X-ray soft $\gamma$-ray Background (shaded area). The three SSC curves
corresponds to different $\nu_{peak}$ values (log $\nu_{peak}= 12.8, 13.5 $ and
$ 13.8$), constrained as described in the text.}
\label{fig:method}       
\end{figure}

The integrated Blazar intensity at microwave frequencies has been computed by
using an updated radio LogN-LogS and it has been extrapolated to the hard
X-rays and soft $\gamma$-rays by using simple SSC models for the SEDs
\cite{gio06}. Figure \ref{fig:method} shows the CMB, CXB and CGB observed
levels, depicted as simple solid lines, together with three SEDs from a simple
homogeneous SSC models. The SED parameters are constrained to
\begin{itemize}
\item be consistent with the expected integrated flux at 94 GHz,
\item have the \amx slope equal to the mean value of the WMAP Blazars (\amx =
1.07),
\item possess a radio spectral slope  equal to the average value of the WMAP
microwave selected Blazars.
\end{itemize}
The three curves, forced to pass through the
three star symbols graphically representing the three constraints listed above,
are characterized by three different synchrotron $\nu_{peak}$ values.

From Fig. \ref{fig:method} we see that a high value of \nupeak over-predicts by
a large factor the observed hard-X-ray to soft $\gamma$-ray
Cosmic Background, whereas a too low value of \nupeak predicts a negligible
contribution. The case Log (\nupeak) = 13.5 Hz predicts 100\% of the
Hard-Xray/Soft $\gamma$-ray Cosmic Background. Since the Log(\nupeak) values of
Blazars in the 1Jy-ARN survey and WMAP catalog peak near 13.5 and range from
12.8 to 13.7 within one sigma from the mean value, the data presently available
indicate that Blazars may be responsible for a large fraction, possibly 100\%
of the Hard-Xray/Soft $\gamma$-ray Cosmic Background. \par

\begin{figure}
\centering
\includegraphics[width=0.35\textwidth,angle=-90]{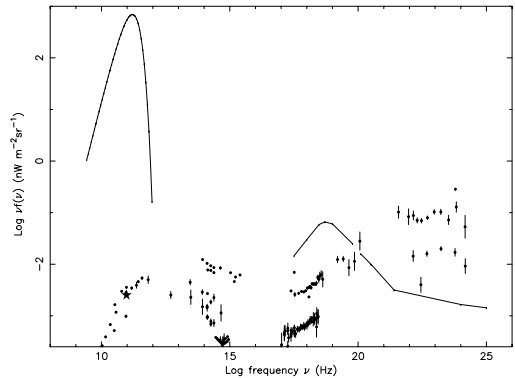}
\caption{The CMB, X-ray and $\gamma$-ray cosmic backgrounds with superimposed the
SED of the Blazar 3C279 scaled as described in the text.}
\label{fig:3c279}       
\end{figure}

Blazars are the large majority of the extragalactic $\gamma$-ray (E$>$ 100 MeV)
identified sources detected by the EGRET experiment. 
In order to estimate Blazar contribution to the $\gamma$-ray Cosmic
Backgrounds, one can analogously scale the full SED of EGRET detected LBL
Blazars, such as that of the well known blazar 3C279, to the integrated Blazar
flux intensity at CMB energies. In Fig.~\ref{fig:3c279} we show the SED of
3C279 scaled so that its flux at 94 GHz matches the cumulative emission of the
entire Blazar population (star symbol).

From Fig. \ref{fig:3c279} one can see that while at X-ray frequencies the
contribution to the CXB ranges from a few \% to over 10\% in the higher states,
the predicted flux at $\gamma$-ray frequencies ranges from about 100\% to
several times the observed Cosmic Background intensity. This large excess
implies that either 3C279 is highly non representative of the class of Blazars,
despite the contribution to the CXB is consistent with other estimates, or its
duty cycle at $\gamma$-ray 
frequencies is very low. The
same approach can be followed with other Blazars detected at $\gamma$-ray
frequencies. In all EGRET detected WMAP Blazars the SED of LBL Blazars
over-predicts the CGB by a large factor.

We define a microwave to $\gamma$-ray slope as
\begin{equation}
 \displaystyle {\alpha_{\mu \gamma} = -{Log(f_{94GHz}/f_{100MeV})\over{Log(\nu_{94GHz}/\nu_{100MeV})}}}~,
 \label{eq.alphamugamma}
\end{equation}
and a limiting value: ${\alpha_{\mu \gamma}}_{100\% CGB} = 0.994$ which is the
value of an hypotetical source that would produce 100\% of the CGB if
representative of the class.\\
Any source with $\alpha_{\mu \gamma} < $ 0.994 should have a duty cycle lower
than 100\% in order not to overproduce the extragalactic diffuse $\gamma$-ray
background.\\
We estimate the blazar duty cycle by defining
\begin{equation}
 \displaystyle {Duty ~ Cycle =100 \times 10^{-11.41 ~(0.994-\alpha_{\mu \gamma})}}~,
 \label{eq.dc}
\end{equation}
where $Log(\nu_{94GHz}/\nu_{100MeV}) = 11.41$.

In the following section we present the preliminary results of our analisis on
a sample of WMAP detected Blazars and for the nominal sensitivity values AGILE
and GLAST instruments we estimate the maximum duty cycle allowed, taking into
account the constraints from the observed extragalactic $\gamma$-ray
background, in order to be detectable and we identify a subset of sources which
can in principle be detectable also in a steady state without over-predicting
the background.

%
\section{Preliminary Results and Discussion}
\label{sec:results}
The subsample we analysed is made of 39 LBL sources,
belonging to the 1st year WMAP bright source catalog, selected at high latitude
$|b|> 30 \deg$ and with measured flux values at 94 GHz.
The general threshold condition to detect a source flux is:
$(signal) \geq n ~ \sigma$, where $\sigma=noise$, and
the signal is equal to $T-B$ (total-background). 
From error propagation one gets $\sigma= \sqrt{T+B}$, from which  
is possible to derive a general ``handy'' sensitivity formula \cite{pittori06} 
reported below to evaluate AGILE and GLAST sensitivities:
\beq
&& S(E_i) = \\ \nonumber
&& \frac{n^2 + \sqrt{n^4 + 8 n^2  F_{diff}                        
A_{eff} ~ T ~ 2 \pi \Bigl( 1- cos~\theta_{PSF}^{68\%} \Bigr)}}
{2 f A_{eff} ~ T ~ \Delta E_i}
\label{final}
\eeq
Parameters characterising the instruments are the effective area
and the PSF 68\% containement radius, that corresponds to the acceptance
solid angle value for diffuse background evaluation.
The corresponding fraction of accepted signal photons is f=0.68.
Note that an $E_i$ dependence of these quantities in the formula is to be understood.
We set n=5, corresponding to a threshold condition on the signal of 5$\sigma$,
we take $\Delta E_i \sim E_i$ and assume any other
efficiency factors =1 to give an estimate of limiting sensitivity values for both experiments.
Values for AGILE/GRID come from \cite{AGILE} and values for GLAST/LAT 
come from \cite{GLAST}.

We evaluated AGILE sensitivity at high latitude for two exposure times:
T = $10^6 s \sim$ 2 weeks which corresponds to a typical AGILE pointing,
and for the nominal lifetime of the mission: T = 2 yrs.
GLAST high latitude sensitivity is evaluated for T = 1 yr.

\begin{figure}
\centering
\includegraphics[width=0.55\textwidth]{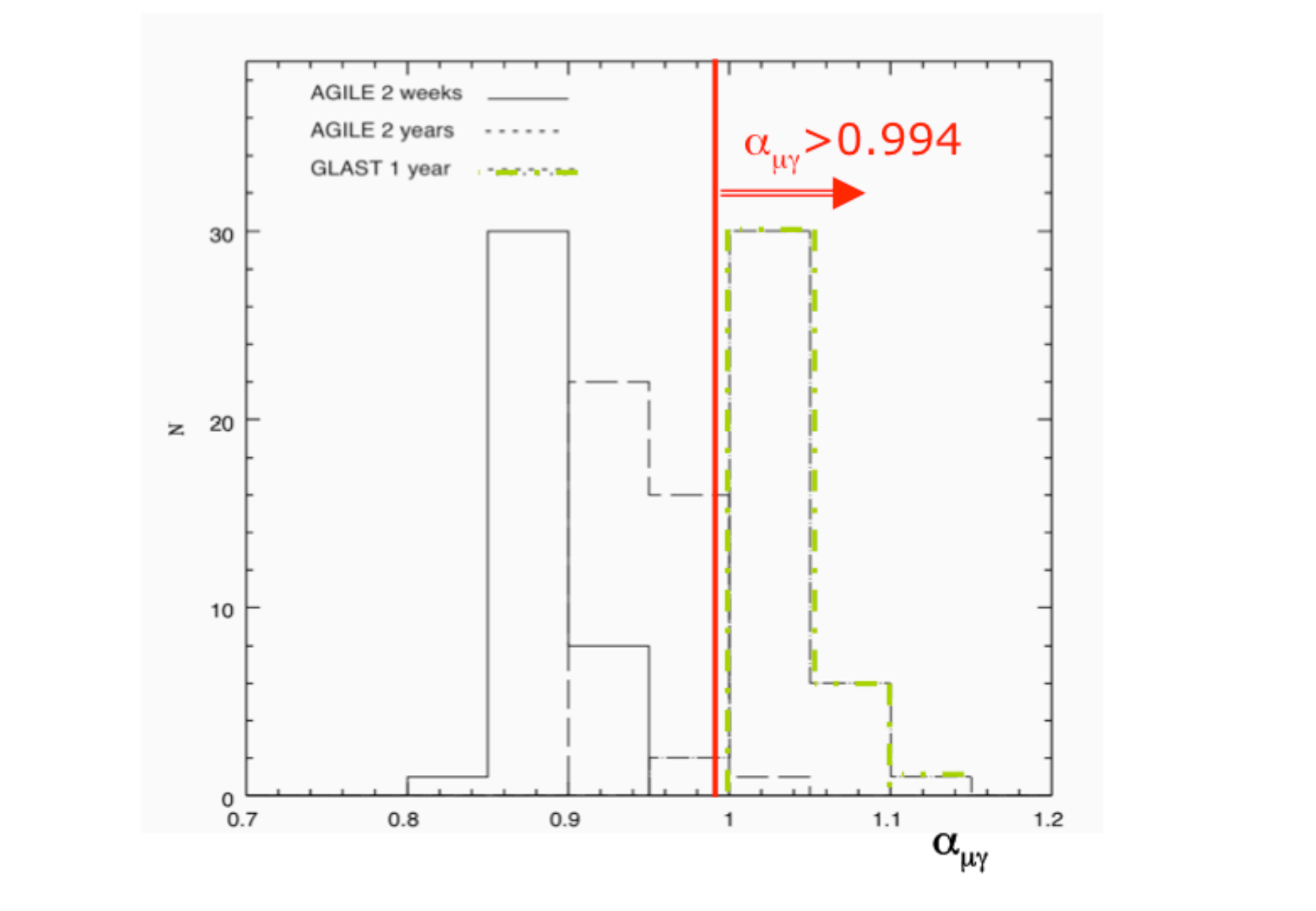}
\caption{$\alpha_{\mu \gamma}$ histogram for a 2 weeks AGILE pointing, the whole 2 years 
AGILE sensitivity and the 1 year GLAST sensitivity. $\alpha_{\mu \gamma} >$ 0.994 corresponds 
to a duty cycle $>$ 100\% that is no background constraints on the observing capability.}
\label{fig:alpha_mugamma_histo}       
\end{figure}

\begin{figure}
\centering
\includegraphics[width=0.55\textwidth]{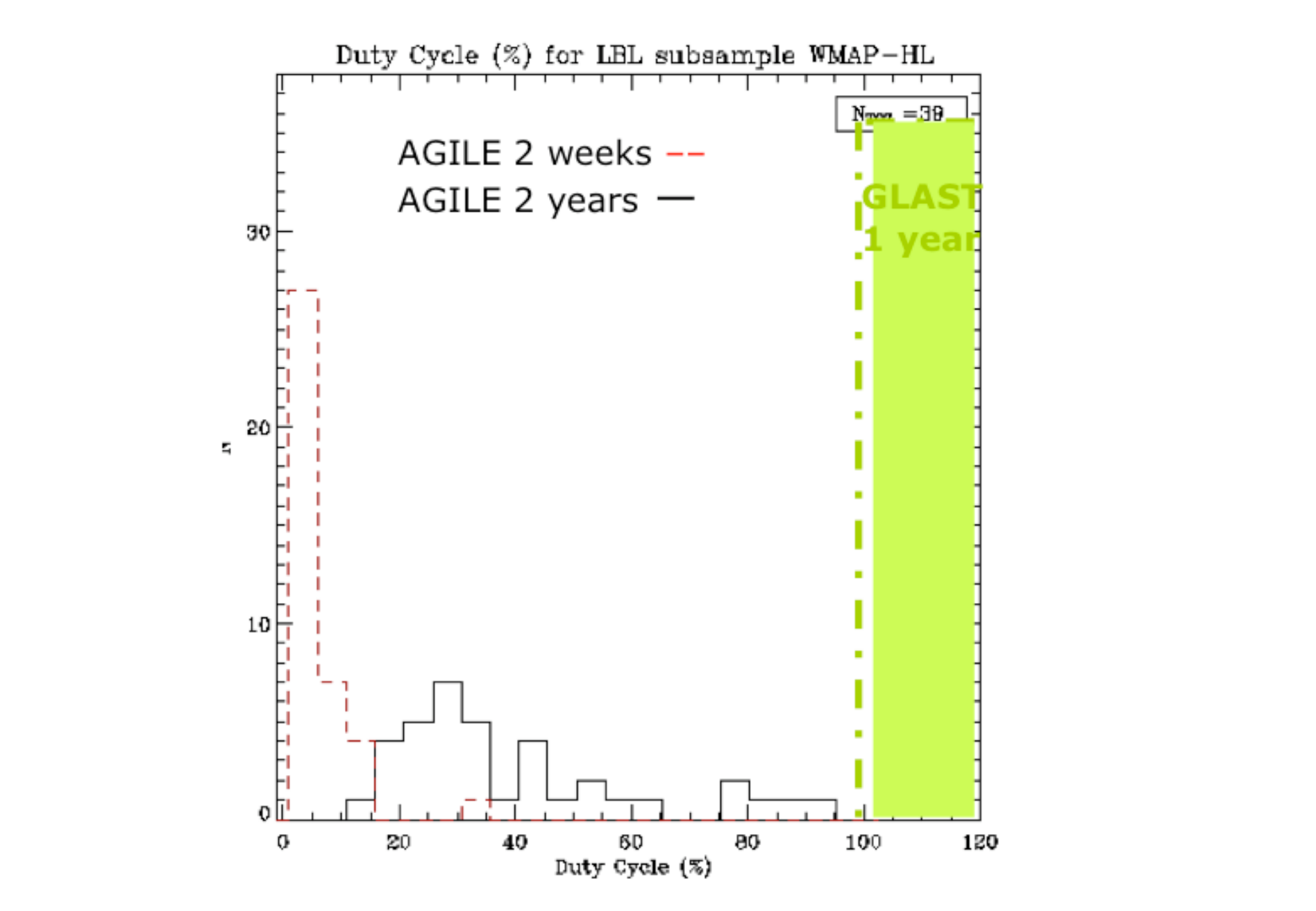}
\caption{Duty cycle distribution for the 39 LBL sources of our subsample. 
GLAST in 1 year
will have no background constraints in observing them.}
\label{fig:dc_histo}       
\end{figure}

In Table~\ref{tab.all} we show our preliminary results for all the 39 sources
in the sample.
Figs.~\ref{fig:alpha_mugamma_histo} and \ref{fig:dc_histo} show the 
distribution of the $\alpha_{\mu\gamma}$ and source duty cycle values
obtained by using AGI\-LE and GLAST sensitivities.

We note that GLAST in one year would be able to detect all High Latitude WMAP sources
in the sample, also in a low-flux steady state with no background constraints.
AGILE in two years would be able to detect a few High Latitude WMAP sources with no
duty cycle constraints such as 3C279, 3C273 and all other sources in the sample
with duty cycle greater than $\sim$ 20\%.
AGILE 2 weeks pointing would detect sources in a flaring state with duty cycle
in the range $\sim$ 1 - 15 \%.

\begin{table*}
\centering
\includegraphics[width=0.45\textwidth,angle=-90]{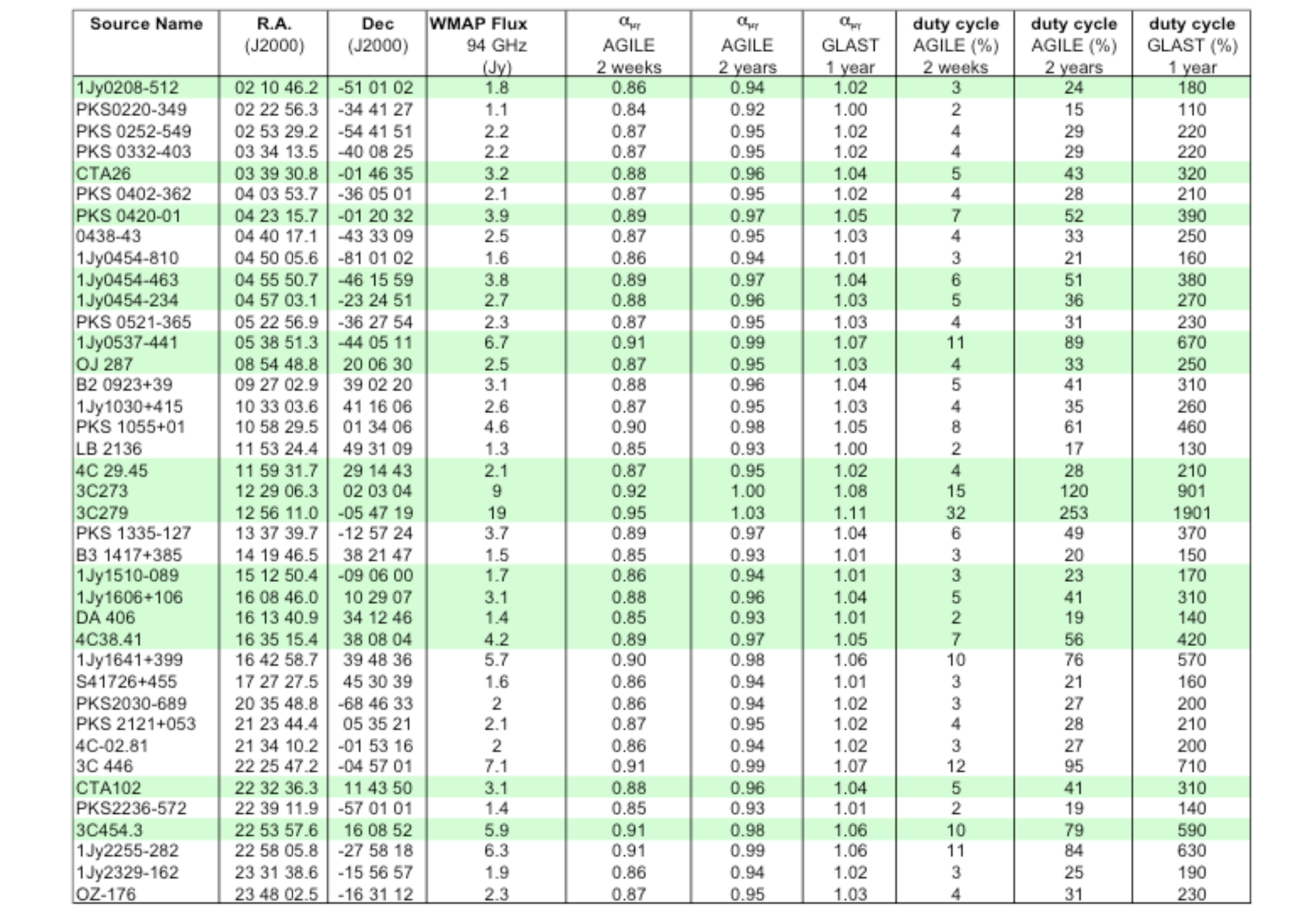}
\caption{Preliminary results for the sample of 39 Low Peaked 
Blazars with WMAP detection, selected at high 
latitude and with measured flux values at 94 GHz. Green-shadowed sources are 
those appearing also in 3EG (subsample of 16 sources).}
\label{tab.all}       
\end{table*}

\begin{figure}
\vskip -0.4truecm
\centering
\includegraphics[width=0.52\textwidth]{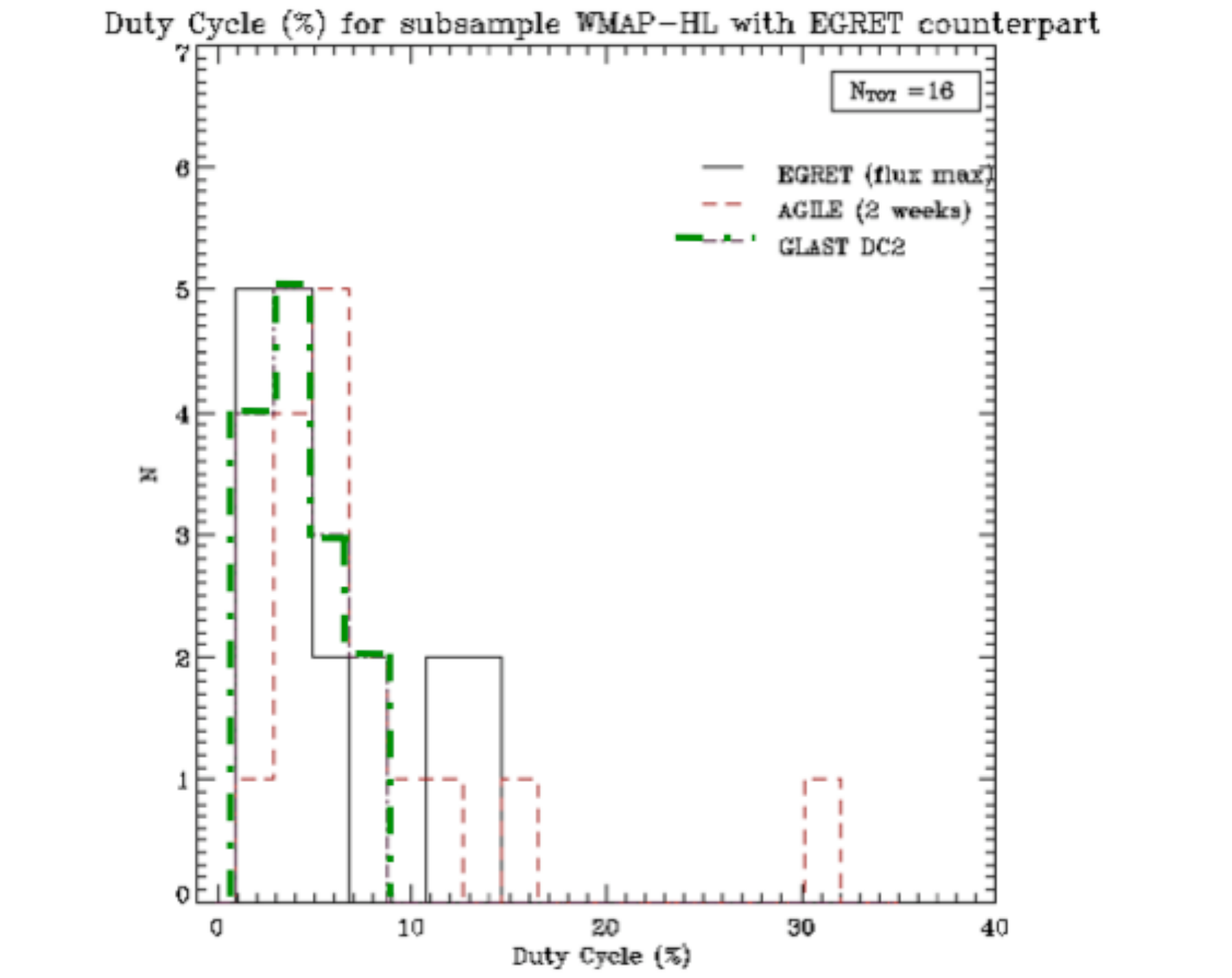}
\caption{Duty cycle distribution for the sub-sample of sources with 
EGRET counterpart:
maximum $\gamma$-ray EGRET flux (solid), AGILE 
sensitivity of one typical pointing (dashed) and GLAST DC2 
simulated flux values (dash-dot).}
\label{fig:dc_egret_histo}       
\end{figure}

In Tables.~\ref{tab:egret_sub1} and ~\ref{tab:egret_sub2} we show the results 
for the subset of WMAP High Latitude sources which also 
appear in the third EGRET catalog (3EG).
We also compare our results with the
GLAST Data Challenge 2 (DC2), 
corresponding to 55 days of realistic simulated $\gamma$-ray data. 
\begin{table}
\centering
\includegraphics[width=0.50\textwidth]{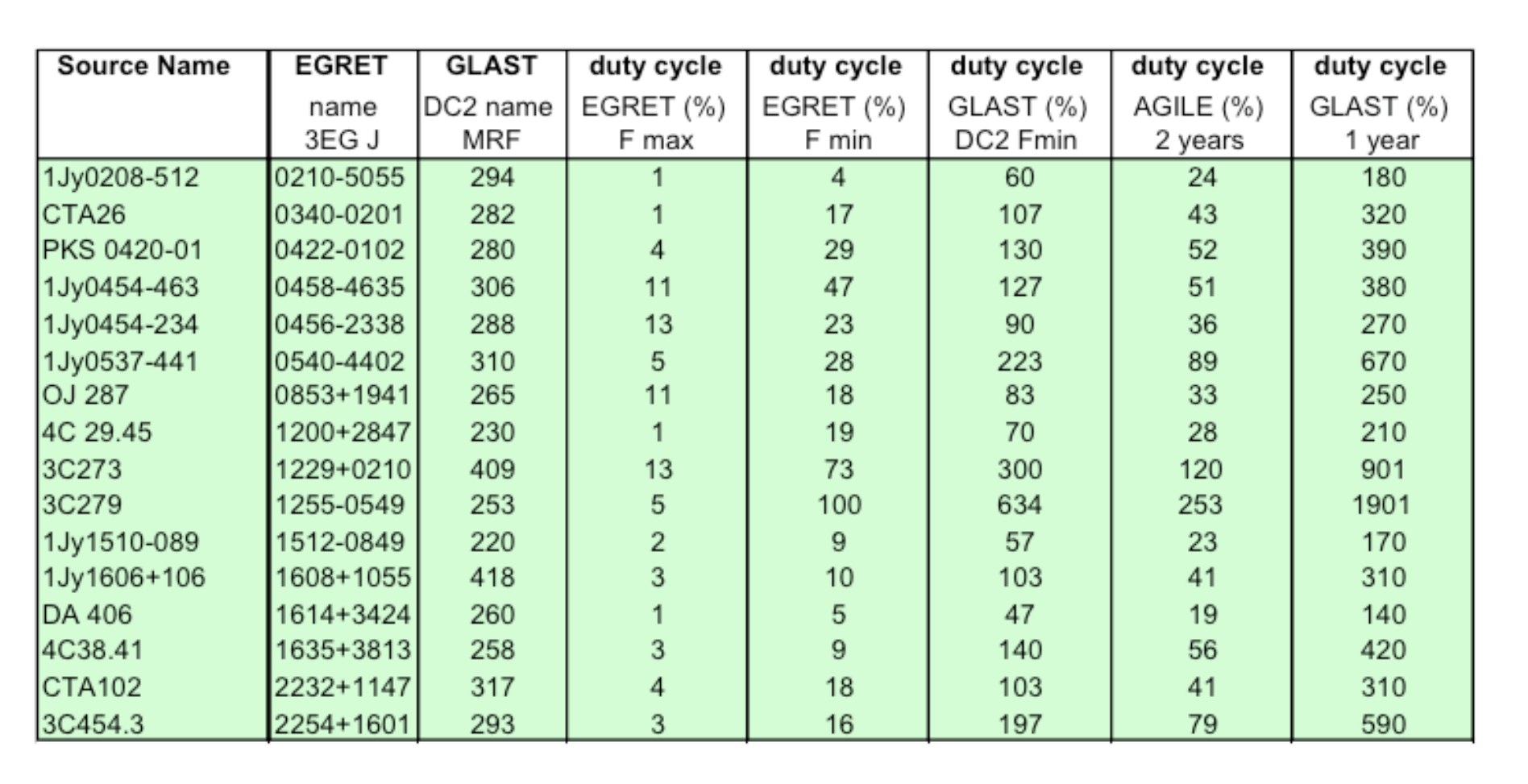}
\caption{Results for subsample of sources in 3EG: 
comparison among max and min observed EGRET fluxes and limiting DC2, 
AGILE 2-years and GLAST 1-year sensitivities.}
\label{tab:egret_sub1}       
\end{table}
%
%
\begin{table}
\centering
\includegraphics[width=0.50\textwidth]{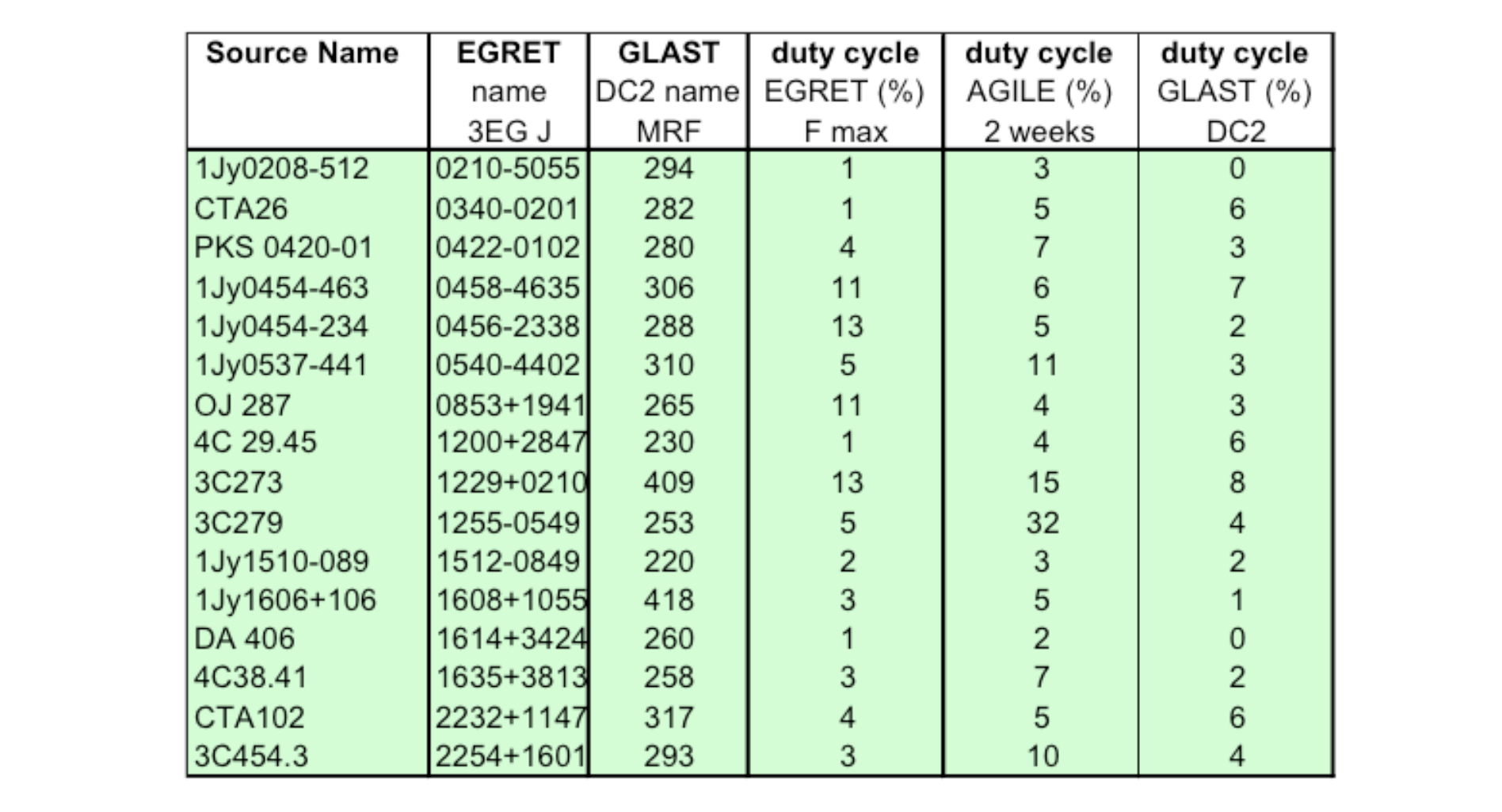}
\caption{Results for subsample of sources in 3EG: 
comparison among max observed EGRET fluxes, AGILE 2-week sensitivity and 
simulated GLAST DC2 fluxes.}
\label{tab:egret_sub2}       
\end{table}
We finally show in Fig.~\ref{fig:dc_egret_histo} the comparison among the Duty
Cycle of the High Latitude WMAP subsample with EGRET counterparts.
We note that both EGRET and DC2 simulated data correspond to sources with high $\gamma$-ray
flux levels, with low duty cycle allowed (in the range 1\% - 15\%) in order not
to overproduce the extragalactic background, assuming that each source is
representative of the entire LBL blazar population. Sources in such high state
could also be detected by AGILE in just one pointing ($\sim$15 days).
%

%
%




\end{document}